\documentclass[twocolumn,superscriptaddress,notitlepage,showpacs,aps,longbibliography,prd]{revtex4-1}

\usepackage{epsf}
\usepackage{amsmath}
\usepackage{amsfonts}
\usepackage{amssymb}
\usepackage{bm}
\usepackage{bbm}
\usepackage{nicefrac}
\usepackage{xcolor}
\usepackage{pifont}
\usepackage{graphicx}
\usepackage{enumerate}
\usepackage{dcolumn}
\usepackage{maybemath}
\usepackage{xfrac}
\usepackage{siunitx}
\usepackage{soul}

\usepackage{color}

\newcommand{\xy}{{\parallel}}

\newcommand{\plus}{{\mbox{\scriptsize $(+)$}}}
\newcommand{\minus}{{\mbox{\scriptsize $(-)$}}}

\newcommand{\magnetron}{{\mathrm{m}}}

\newcommand{\hate}{\hat{\mathrm e}}

\newcommand{\ii}{\mathrm{i}}

\newcommand{\qq}{\ell}

\newcommand{\PT}{{\mathrm{T}}}

\newcommand{\QED}{{\mathrm{QED}}}

\newcommand{\calE}{\mathcal{E}}
\newcommand{\calO}{\mathcal{O}}

\newcommand{\calH}{\mathcal{H}}

\definecolor{garrosgreen}{rgb}{0.1, 0.4, 0.1}
\definecolor{dartmouthgreen}{rgb}{0.05, 0.5, 0.06}
\definecolor{duelferred}{rgb}{0.7, 0.2, 0.1}
\definecolor{cambridgeblue}{rgb}{0.1, 0.3, 1.0}
\definecolor{oxfordblue}{rgb}{0.05, 0.2, 0.7}

\definecolor{irishgreen}{rgb}{0.1, 0.7, 0.3}

\begin{document}

\title{Foldy--Wouthuysen Transformation
in Strong Magnetic Fields\\
and Relativistic Corrections for Quantum Cyclotron Energy Levels}

\author{Albert Wienczek}
\affiliation{Department of Physics and LAMOR,
Missouri University of Science and Technology,
Rolla, Missouri 65409, USA}
\affiliation{Faculty of Physics, University of 
Warsaw, Pasteura 5, 02--093 Warsaw, Poland}

\author{Christopher Moore} 
\affiliation{Department of Physics and LAMOR,
Missouri University of Science and Technology,
Rolla, Missouri 65409, USA}

\author{Ulrich D. Jentschura}
\email{email: ulj@mst.edu}
\affiliation{Department of Physics and LAMOR,
Missouri University of Science and Technology,
Rolla, Missouri 65409, USA}

\begin{abstract}
We carry out a direct, iterative Foldy--Wouthuysen
transformation of a general Dirac Hamiltonian coupled 
to an electromagnetic field, including the anomalous
magnetic moment.
The transformation is carried out 
through an iterative disentangling of the particle and
antiparticle Hamiltonians, in the expansion for higher orders
of the momenta.
The time-derivative term from the unitary transformation 
is found to be crucial in supplementing the transverse component of the 
electric field in higher orders.
Final expressions are obtained for general combined electric
and magnetic fields, including strong magnetic fields. 
The time-derivative of the electric field 
is shown to enter only in the seventh 
order of the fine-structure
constant if the transformation is carried out in the standard
fashion. We put special emphasis on the case of strong fields,
which are important for a number of applications,
such as electrons bound in Penning traps.
\end{abstract}

\maketitle

\small

\tableofcontents

\normalsize

%
% Introduction
%
\section{Introduction}
\label{sec1}

The purpose of this paper is threefold:
First, to discuss the role of nonstandard 
and standard Foldy--Wouthuysen transformations, 
up to seventh order in the momenta, 
second, to clarify the role 
of higher-order terms in the magnetic fields
which become relevant for particles bound
in strong magnetic fields (Penning traps),
and third, to apply the results to 
the calculation of quantum cyclotron energy 
levels.

Let us start with the first purpose, which requires
some background discussion.
The Foldy--Wouthuysen transformation~\cite{FoWu1950} 
is a cornerstone of the description of electronic bound states in 
simple atomic systems.
The purpose of the transformation is to start from
a (generalized) Dirac Hamiltonian,
and to disentangle the particle and antiparticle
degrees of freedom.
We recall that the Dirac Hamiltonian describes 
particles and antiparticles simultaneously,
and the Foldy--Wouthuysen transformation
is used to find separate effective Hamiltonians for the 
particle (positive-energy) and antiparticle (negative-energy) states.
In simple cases, such as a free electron, 
one can disentangle the particle and antiparticle 
Hamiltonians to all orders in
the coupling parameter~\cite{BjDr1964},
but this is, in general, not possible when the Dirac 
particle is bound in external fields, because of difficulties 
in expressing infinite series of multi-commutators 
in closed analytic form.
One can do the exact transformation
(to all orders in the momenta) only in rare cases.
As a consequence, for atomic bound states, 
one resorts to a perturbative
scheme, which involves an expansion in higher
orders of the momenta, or in powers of a suitably chosen
coupling parameter. The coupling parameter can be
the fine-structure constant $\alpha = \alpha_\QED \approx 
1/137.036$
or a suitable generalization
(see Ref.~\cite{Je2014pra}).

For an electron bound to a nucleus, 
in the fourth order in the momenta,
starting from the Dirac--Coulomb Hamiltonian,
one obtains~\cite{BjDr1964,ItZu1980} the relativistic corrections
to the hydrogen bound states, i.e.,
the relativistic $p^4$ correction,
the zitterbewegung term (the Darwin term),
and the spin-orbit coupling (Russell--Saunders coupling).
The Foldy--Wouthuysen method was
generalized to sixth order in the momenta in 
Ref.~\cite{Pa2005}, using a non-standard
transformation given in Eq.~(8) of Ref.~\cite{Pa2005},
which makes the decoupling transformation
computationally easier. It
gives rise to a term [see Eq.~(18) of Ref.~\cite{Pa2005}]
which involves the time-derivative of the electric field,
\begin{equation}
H \sim -\frac{e}{16 m^3} 
\{
\sigma \cdot \vec \pi, \,
\sigma \cdot \dot{\vec E} \} \,,
\end{equation}
where $\vec\pi = \vec p - e \, \vec A$ is the 
kinetic momentum 
(throughout this paper, $e= -|e|$ is the electron charge). 
Furthermore,
$\vec E$ is the electric field, and 
$\{ A, B \} = A \, B + B \, A$ denotes the anticommutator.
One may eliminate 
the time-derivative of the electric field by an 
additional unitary transformation given in Eq.~(19) 
of Ref.~\cite{PaYePa2016}.

Throughout this paper, we use Coulomb gauge, so that 
one can easily identify the longitudinal and 
transverse parts of the electric field $\vec E$,
which are related to the vector potential $\vec A$ as
\begin{equation}
\label{EperpEpar}
\vec A = \vec A_\perp \,, \qquad
\vec E_\parallel = - \vec\nabla A^0 \,, \qquad
\vec E_\perp = - \frac{\partial}{\partial t} \vec A \,,
\end{equation}
where $\perp$ denotes the transverse (divergence-free) 
field, and $\parallel$ denotes the longitudinal
(curl-free) field component.
Furthermore, we shall use the convention
\begin{equation}
\label{defV}
V = e \, A^0 
\end{equation}
for the binding potential (here, $e$ is the electron charge).
In the treatment of atomic bound states,
the binding potential is often approximated by the 
Coulomb potential $V(r) = -Z\alpha/r$,
where $Z$ is the nuclear charge number,
$\alpha$ is the fine-structure constant,
and $r$ is the distance of the orbiting particle 
and the nucleus. However, in the treatment of 
bound states in a Penning trap~\cite{BrGa1982,Br1985aop,BrGa1986},
the binding potential is given by the electric quadrupole 
field of the Penning trap, while an additional 
strong magnetic field provides the 
axial confinement. The binding Coulomb field 
is replaced by the binding field of the Penning trap.

The problem of the calculation 
of the higher-order corrections to the 
Foldy--Wouthuysen transformation 
has been considered, quite recently, in Ref.~\cite{ZaPa2010},
where the general Hamiltonian for particles with 
arbitrary spin has been investigated. 
The special case of $s = 1/2$ has been treated
in Eqs.~(36)--(38) of Ref.~\cite{ZaPa2010}.  
Furthermore, in Eq. (7) of Ref.~\cite{HaZhKoKa2020},
a Hamiltonian has been indicated which has been obtained from 
the nonrelativistic quantum electrodynamics (NRQED)
approach outlined in Ref.~\cite{HiLePaSo2013}.
Indeed, in Ref.~\cite{HiLePaSo2013},
the coefficients in the effective Hamiltonian
were obtained by matching of the NRQED Hamiltonian,
given in Eq.~(1) of Ref.~\cite{HiLePaSo2013},
with scattering amplitude calculations.
It is an interesting question to compare the NRQED approach to the 
sixth-order generalization of the standard Foldy--Wouthuysen
transformation, outlined in Refs.~\cite{FoWu1950,BjDr1964,ItZu1980}.

Thus, we here present an application of the 
{\em standard} Foldy--Wouthuysen 
transformation~\cite{FoWu1950,BjDr1964,ItZu1980},
which is based on the 
iterative elimination of ``odd'' operators in the
Hamiltonian via unitary transformations,
to sixth order in the momenta,
for general electric and magnetic fields.
Our approach is sufficiently general to be valid for strong 
magnetic fields, and is thus applicable
to electrons bound to Penning traps~\cite{BrGa1982,Br1985aop,BrGa1986}.
The bound states in Penning traps differ from atomic 
bound states in the sense that 
the primary binding fields are the static 
magnetic field, directed along the trap axis,
and the electric quadrupole field of the trap.
The standard Foldy--Wouthuysen approach, in higher
orders, offers technical difficulties which 
are overcome in the current investigation.

In particular, we do not rely on any 
non-standard transformations, which
were otherwise used in Ref.~\cite{Pa2005}.
As a consequence, we are able to compare the 
standard sixth-order Foldy--Wouthuysen approach
to the generalized nonstandard Foldy--Wouthuysen transformations
outlined in Ref.~\cite{Pa2005}.
In the standard approach, 
one eliminates odd operators 
by a well-defined iterative procedure.
Here, the odd operators are 
understood as the off-diagonal entries
in the bispinor basis. Let us consider an example.
For a Hermitian 
Hamiltonian $H$, of the form
\begin{equation}
H = \left( \begin{array}{cc}
\calE & \calO \\
\calO^\dagger & \calE' \\
\end{array} \right) \,,
\qquad
H^\dagger = H \,,
\end{equation}
where $\calE = \calE^\dagger$, $\calE' = \calE'^\dagger$ 
and $\calO$ are $2 \times 2$ matrices,
the odd operator is just $\calO$. The iterative elimination
of $\calO$, through successive applications of the unitary 
transformations, is the aim of the Foldy--Wouthuysen method.

Units with $\hbar = c = \epsilon_0 = 1$ are employed.
This paper is organized as follows.
In Sec.~\ref{sec2}, we consider the scaling 
of operators for an electron bound in a Penning trap.
This scaling is different from that encountered in 
an atom, because the binding fields (in particular, 
the magnetic field) have to be given more weight.
In particular, the magnetic field enters at a lower
order in a generalized coupling parameter
(generalized fine-structure constant) than in 
atomic bound systems. In fact, in Sec.~\ref{sec2}, 
we define suitable generalized coupling parameters for the 
electron bound in the Penning trap. 
In Sec.~\ref{sec3}, we carry out the main part of the 
calculations for the sixth-order, and seventh-order,
Foldy--Wouthuysen transformation pertaining to 
bound electrons. 
We obtain general results which allow us to 
carry out a detailed comparison between the 
different approaches previously pursued in 
Refs.~\cite{Pa2005,ZaPa2010,PaYePa2016,HiLePaSo2013,HaZhKoKa2020}.
We then specialize the general expressions to the 
case of a Penning trap (Sec.~\ref{sec4}), 
and derive a few higher-order terms,
supplementing previous investigations~\cite{BrGa1982,Br1985aop,BrGa1986}.
These are important for the determination
of the fine-structure constant from measurements
of the anomalous magnetic moment of the electron.
Finally, in Sec.~\ref{sec5}, we draw some conclusions.

%
% Preparatory Considerations
%
\section{Preparatory Considerations}
\label{sec2}

%
% Penning Trap
%
\subsection{Penning Trap}
\label{sec21}

We shall attempt to devise a formalism for the 
systematic analysis of higher-order corrections to 
the bound energy levels for electrons 
bound in field configurations where the
wave functions are spatially confined by the field 
geometry, and a discrete spectrum of bound states
results.
Due to the spatial confinement, 
one obtains a discrete spectrum of bound states.
Our investigations are motivated, to a large extent,
by the necessity to extend the usual Foldy--Wouthuysen
formalism to situations with strong confining magnetic 
fields. An example is given by an electron confined 
in a Penning trap (see 
Refs.~\cite{BrGa1982,Br1985aop,BrGa1986}).

In a Penning trap, one has a 
strong, constant, uniform,
confining, magnetic field along the trap axis 
(the $z$ axis), given as $\vec B_\PT = \hate_z \, B_\PT$.
The corresponding vector potential is
\begin{equation}
\vec A_\PT 
= \frac12 \, ( \vec B_\PT \times \vec r ) 
= \frac12 \, ( \vec B_\PT \times \vec \rho ) \,,
\end{equation}
where $\vec\rho$ is the position vector in the 
$xy$ plane,
\begin{equation}
\vec \rho = \vec r_\parallel = 
x \, \hate_x + y \, \hate_y \,.
\end{equation}
We decompose the momentum operator as
$\vec p = \vec p_\xy + \vec p_\perp$,
where $\vec p_\xy = p_x \hate_x + p_y \, \hate_y$,
and $\vec p_\perp = p_z \hate_z$.
The kinetic trap momentum $\vec\pi_\PT$ is
\begin{align}
\label{defpiT}
\vec\pi_\PT =& \; \vec p - e \vec A_\PT 
= \vec p_\xy - \frac{e}{2} \, ( \vec B_\PT \times \vec r ) 
+ \vec p_\perp
= \vec \pi_\xy + \vec p_\perp \,,
\\[0.1133ex]
\vec \pi_\xy =& \; \vec p_\xy
- \frac{e}{2} \, ( \vec B_\PT \times \vec r ) \,.
\qquad
\end{align}
The 
scalar potential is $A^0$.
The quadrupole potential $V$ of the Penning 
trap is
\begin{subequations}
\label{potV}
\begin{align}
V =& \; e \, A^0 = V_0 \, \frac{z^2 - \tfrac12 \rho^2}{2 d^2} 
= V_z + V_\xy \,,
\\[0.1133ex]
\label{Vpar}
V_z =& \; \frac12 m \omega_z^2 z^2 
\qquad
V_\xy = - \frac14 m \omega_z^2 \rho^2 \,,
\\[0.1133ex]
\omega_z^2 =& \; \frac{V_0}{m d^2} \,,
\end{align}
\end{subequations}
where $V_0 > 0$ and $d > 0$ are constants. 
Note that $V_\xy$ is repulsive, while
$V_z$ is an attractive harmonic potential.
We also note that
$V_0$ has physical dimension of energy,
and $d$ has a physical dimension of length.
We found it convenient to absorb
the elementary charge $e$
in the definition of $V_0$, which leads to 
a slight change in the notation as compared 
to Ref.~\cite{BrGa1986}. 
We write the spin $g$ factor of the 
electron as $g = 2 (1 + \kappa)$, where $\kappa \approx
\alpha/(2\pi)$ is the anomalous magnetic-moment term~\cite{Sc1948}.
The nonrelativistic Hamiltonian, including the 
anomalous magnetic-moment term, can be written as
\begin{align}
\label{H0rep1}
H_0 =& \; \frac{( \vec\sigma \cdot \vec \pi_\PT)^2 }{2 m} + V
- \frac{e}{2m} \, \kappa \, \vec\sigma\cdot \vec B_\PT 
\nonumber\\[0.1133ex]
=& \; \frac{( \vec\sigma \cdot \vec \pi_\xy)^2 }{2 m} 
- \frac{e}{2m} \, \kappa \, \vec\sigma\cdot \vec B_\PT 
+ \frac{p_z^2}{2 m} + V \,.
\end{align}
Using the result
\begin{align}
( \vec\sigma \cdot \vec \pi_\xy)^2 =& \;
%%% \vec \pi_\xy^2 - e \vec\sigma\cdot \vec B_\PT = 
\vec p_\xy^{\,2} - e \vec L \cdot \vec B_\PT 
+ \frac{m^2 \omega_c^2}{4} \,\rho^2 
- e \vec\sigma\cdot \vec B_\PT \,,
\\[0.1133ex]
\label{defomegac}
\omega_c =& \; \frac{|e| \, B_\PT}{m} \,,
\end{align}
where $\omega_c$ is the cyclotron frequency
and $\vec L$ is the angular momentum operator,
one can write $H_0 = H_\xy + H_\sigma + H_z$
as the sum of an
orbital Hamiltonian $H_\parallel$ which acts in the 
$xy$ plane, of a magnetic Hamiltonian $H_\sigma$ 
which couples to the spin, 
and of a Hamiltonian $H_z$ which 
confines the particle along the $z$ axis,
in a harmonic potential due to the quadrupole 
field of the trap,
\begin{subequations}
\label{H0rep2}
\begin{align}
H_0 =& \; H_\xy + H_\sigma + H_z \,,
\\[0.1133ex]
\label{H0rep2b}
H_\xy =& \; \frac{\vec p_\xy^{\,2}}{2m} 
- \frac{e}{2m} \vec L \cdot \vec B_\PT 
+ \frac{m \omega_c^2}{8} \,\rho^2 + V_\xy \,,
\\[0.1133ex]
H_\sigma =& \; - \frac{e}{2m} \,
(1 + \kappa) \, \vec\sigma\cdot \vec B_\PT \,,
\\[0.1133ex]
H_z =& \; \frac{p_z^2}{2 m} + V_z \,.
\end{align}
\end{subequations}
Due to its harmonic-oscillator structure,
$H_z$ can be written as
\begin{equation}
H_z = \frac{p_z^2}{2 m} + \frac12 m \omega_z^2 z^2 
= \omega_z \, \left( a_z^\dagger \, a_z + \frac12 \right) \,.
\end{equation}
where the lowering and raising operators
$a_z$ and $a_z^\dagger$ are the usual ones 
for a quantum harmonic oscillator,
\begin{subequations}
\begin{align}
a_z =& \; \sqrt{ \frac{m \omega_z}{2} } \, z +
\ii \left( \frac{1}{2 m \omega_z} \right)^{1/2} \, p_z \,,
\\[0.1133ex]
a_z^\dagger =& \; \sqrt{ \frac{m \omega_z}{2} } \, z -
\ii \left( \frac{1}{2 m \omega_z} \right)^{1/2} \, p_z \,.
\end{align}
\end{subequations}
With reference to Eq.~\eqref{H0rep2b}, 
we have the relations
\begin{subequations}
\begin{align}
\label{defHxy}
H_\xy =& \; \frac{\vec p_\parallel^{\,2}}{2 m} +
\frac{\omega_c}{2} L_z +
\frac{m (\omega_c^2 - 2 \omega_z^2)}{8} \, \rho^2
\\[0.1133ex]
= & \;
\omega_\plus \left( a_\plus^\dagger \, a_\plus + \frac12 \right) -
\omega_\minus \left( a_\minus^\dagger \, a_\minus + \frac12 \right) \,,
\nonumber\\[0.1133ex]
\label{defomegaplus}
\omega_\plus =& \; \frac12 \, \left( \omega_c +
\sqrt{ \omega_c^2 - 2 \omega_z^2 } \right) \,,
\\[0.1133ex]
\label{defomegaminus}
\omega_\minus =& \; \frac12 \, \left( \omega_c -
\sqrt{ \omega_c^2 - 2 \omega_z^2 } \right) 
= \omega_\magnetron 
\approx \frac{\omega_z^2}{2 \omega_c} \,.
\end{align}
\end{subequations}
The quantities $\omega_\plus$ and $\omega_\minus$ are
corrected cyclotron ($\omega_c$) and 
magnetron ($\omega_\magnetron$) frequencies;
they correspond to the conventions used 
in Eq.~(2.14) of Ref.~\cite{BrGa1986}.
Physically, we can understand the minus sign in front of the 
$\omega_\minus$ term in Eq.~\eqref{defHxy}
in terms of the repulsive character of
the potential $V_\parallel$ defined in Eq.~\eqref{Vpar}.
The lowering and raising operators
$a_\plus$, $a_\minus$, $a_\plus^\dagger$ and 
$a_\minus^\dagger$ are given in Eqs.~(2.48a) and (2.48b) of Ref.~\cite{BrGa1986}.
The operators $a_\plus$ and $a_\plus^\dagger$ 
are associated with the cyclotron motion,
\begin{align}
a_{\plus} = & \;
\left( \frac{m}{2 \, \left( \omega_{\plus} - \omega_{\minus} \right)} 
\right)^{1/2} \,
\left( V_{\plus x} - \ii \, V_{\plus y} \right) \,,
\\[0.1133ex]
a^\dagger_{\plus} = & \;
\left( \frac{m}{2 \, \left( \omega_{\plus} - \omega_{\minus} \right)} 
\right)^{1/2} \,
\left( V_{\plus x} + \ii \, V_{\plus y} \right) \,,
\end{align}
and we shall consider the operator $\vec V_\plus$, whose $x$
and $y$ components enter the definition of 
$a_{\plus}$ and $a^\dagger_{\plus}$, in the following,
but first, let us consider 
the lowering and raising operators of the magnetron 
motion. One has the lowering operator $a_{\minus}$
and the raising operator $a^\dagger_{\minus}$,
\begin{align}
a_{\minus} = & \;
\left( \frac{m}{2 \, \left( \omega_{\plus} - \omega_{\minus} \right)}
\right)^{1/2} \,
\left( V_{\minus x} + \ii \, V_{\minus y} \right) \,,
\\[0.1133ex]
a^\dagger_{\minus} = & \;
\left( \frac{m}{2 \, \left( \omega_{\plus} - \omega_{\minus} \right)}
\right)^{1/2} \,
\left( V_{\minus x} -\ii \, V_{\minus y} \right) \,,
\end{align}

The quantum-mechanical formulation of the 
vector-valued operators 
$\vec V_\plus$ and
$\vec V_\minus$ can in principle be inferred from the 
quantum-classical correspondence indicated in 
Eqs.~(2.13) and (2.42) of Ref.~\cite{BrGa1986}.
The operators $\vec V_\plus$ and $\vec V_\minus$ are
vector-valued and act in the $xy$ plane.
It is instructive to indicate
the explicit formulas,
\begin{subequations}
\begin{align}
\vec V_\plus =& \;
\frac{\vec p_\parallel}{m} + \frac12 \,
\sqrt{ \omega_c^2 - 2 \, \omega_z^2 } \,
(\hate_z \times \vec \rho) \,,
\\[0.1133ex]
\vec V_\minus =& \;
\frac{\vec p_\parallel}{m} - \frac12 \,
\sqrt{ \omega_c^2 - 2 \, \omega_z^2 } \,
(\hate_z \times \vec \rho) \,.
\end{align}
\end{subequations}
An interesting feature is that the algebra of the 
cyclotron and magnetron lowering and raising 
operators commute,
\begin{equation}
\left[ a_\minus, a_\plus \right] =
\left[ a_\minus, a_\plus^\dagger \right] = 
\left[ a^\dagger_\minus, a_\plus^\dagger \right] = 0 \,.
\end{equation}
This means that we can raise cyclotron and magnetron
quantum numbers independently by using the 
$a_\plus$ and $a_\plus^\dagger$,
and $a_\minus$ and $a_\minus^\dagger$ operators.

\begin{figure}[t!]
\begin{center}
\begin{minipage}{0.99\linewidth}
\begin{center}
\includegraphics[width=0.99\linewidth]{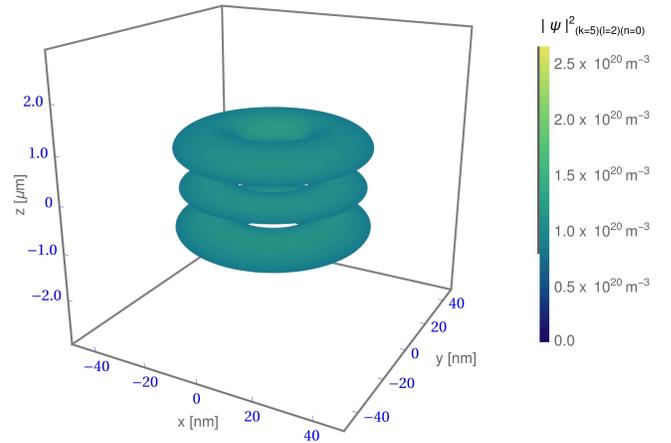}
\caption{\label{fig1} 
We display the 
probability density $| \psi |^2 = | \psi_{k \ell n s}(\vec r) |^2$
of the quantum cyclotron state with 
quantum numbers $k=2$, $n=0$, and $\qq = 2$
[see Eq.~\eqref{psidef}].
This is the second axial excited state ($k=2$), 
the cyclotron ground state ($n = 0$),
and the second excited magnetron state 
($\qq = 2$).
The probability density is independent of the 
spin state ($s = \pm 1$).
We use parameters from Ref.~\cite{BrGa1986},
i.e., $\omega_c = 2 \pi \times 164.4 \, {\rm GHz}$,
$\omega_z = 2 \pi \times 64.42 \, {\rm MHz}$,
which implies that the corrected magnetron frequency 
is $\omega_\minus = 2 \pi \times 12.62 \, {\rm kHz}$.
Axial states with high average excitation 
form the basis of experiments~\cite{FaGa2021prl,FaGa2021pra}.
One notes the large extent of the 
wave function in the axial direction,
which is in the range of micrometers,
while the confining magnetic field of the 
trap restricts the wave function in the 
$x$ and $y$ directions to range of about 50 nanometers.}
\end{center}
\end{minipage}
\end{center}
\end{figure}

\begin{figure}[t!]
\begin{center}
\begin{minipage}{0.99\linewidth}
\begin{center}
\includegraphics[width=0.99\linewidth]{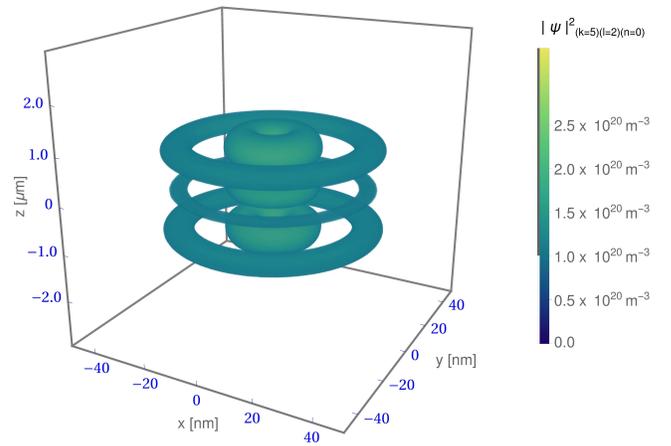}
\caption{\label{fig2}
We present the analogue of Fig.~\ref{fig1}
for the quantum cyclotron state with
quantum numbers $k=2$, $n=1$, and $\qq = 2$.
In contrast to Fig.~\ref{fig1},
this is the first excited cyclotron state ($n = 1$).
Again, we use parameters from Ref.~\cite{BrGa1986},
i.e., $\omega_c = 2 \pi \times 164.4 \, {\rm GHz}$,
and $\omega_z = 2 \pi \times 64.42 \, {\rm MHz}$.
One notes the large extent of the
wave function in the axial direction,
while the wave function is much more confined
in the $x$ and $y$ directions.}
\end{center}
\end{minipage}
\end{center}
\end{figure}

%
% Unperturbed Eigenfunctions
%
\subsection{Unperturbed Eigenfunctions}
\label{sec22}
 
\begin{figure}[t!]
\begin{center}
\begin{minipage}{0.99\linewidth}
\begin{center}
\includegraphics[width=0.99\linewidth]{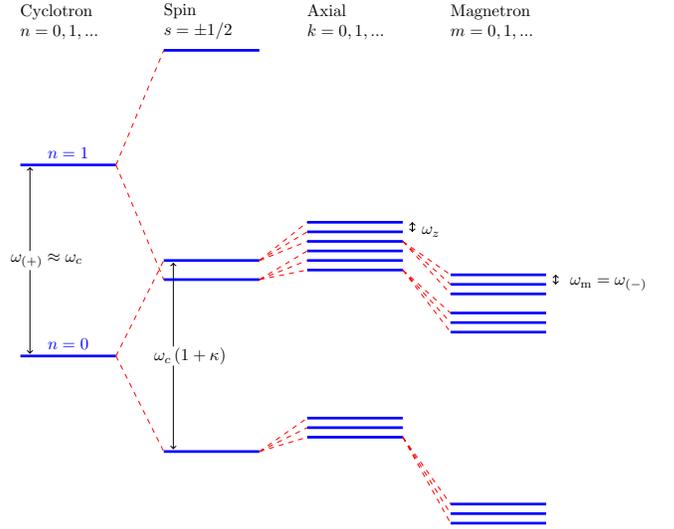}
\caption{\label{fig3} 
The figure provides
an illustration of the bound-state spectrum of an electron 
in a Penning trap.
The cyclotron levels are separated by the frequency 
$\omega_\plus \approx \omega_c$,
while the dominant contribution comes
from the spin projection $s = \pm 1/2$,
because the spin-flip frequency 
is $\omega_c (1 + \kappa)$, where 
$\kappa \approx \alpha/(2 \pi)$. 
Two important quasi-degenerate levels have
the quantum numbers $s = +1/2$, $n = 0$
(spin-up cyclotron ground state),
and $s = -1/2$, and $n = 1$
(spin-down first excited cyclotron state).
They are energetically degenerate 
were it not for the effect of the anomalous magnetic 
moment of the electron.}
\end{center}
\end{minipage}
\end{center}
\end{figure}

Eigenfunctions of the unperturbed, nonrelativistic Hamiltonian
$H_0$ [see Eqs.~\eqref{H0rep1} and~\eqref{H0rep2}] 
are described by the spin projection quantum number $s$, the 
axial quantum number $k$, and the magnetron quantum number $\qq$,
illustrating the fact that an electron bound in a 
Penning trap merely constitutes an ``artificial atom'',
with the trap fields replacing the binding Coulomb field.
The quantum numbers take the following values:
%
%%% \begin{subequations}
\begin{align}
k =& \; 0,1,2,\dots \,, &
\qquad & \mbox{(axial)} \,,
\\[0.1133ex]
\qq =& \; 0,1,2,\dots \,, &
\qquad & \mbox{(magnetron)} \,,
\\[0.1133ex]
n =& \; 0,1,2,\dots &
\qquad & \mbox{(cyclotron)} \,,
\\[0.1133ex]
s =& \; \pm 1 \,,  &
\qquad & \mbox{(spin)} \,.
\end{align}
%%% \end{subequations}
%
The energy eigenvalues of $H_0$ are not bounded from below,
in view of the repulsive character of the radial 
quadrupole potential,
\begin{align}
\label{Esnkq}
E_{k \qq n s} =& \; \omega_c (1+\kappa) \, \frac{s}{2}
+ \omega_\plus \left( n + \frac12 \right)
\nonumber\\[0.1133ex]
& \; + \omega_z \left( k + \frac12 \right)
- \omega_\minus \left( \qq + \frac12 \right) \,.
\end{align}
One takes notice of the negative sign in front of the last
term. The eigenfunctions of the unperturbed Hamiltonian 
can be constructed as (see Figs.~\ref{fig1} and~\ref{fig2})
\begin{align}
\label{psidef}
\psi_{k \qq n s}(\vec r) = & \;
\frac{\left( a^\dagger_z \right)^k}{\sqrt{k!}} \,
\frac{\left( a^\dagger_\minus \right)^\qq}{\sqrt{\qq !}} \,
\frac{\left( a^\dagger_\plus \right)^n}{\sqrt{n!}} \,
\psi_0(\vec r) \, \chi_{s/2} \,,
\\[0.1133ex]
\chi_{1/2} =& \; \left( \begin{array}{c} 1 \\ 0 \end{array} \right) \,,
\qquad
\chi_{-1/2} = \left( \begin{array}{c} 0 \\ 1 \end{array} \right) \,.
\end{align}
where the $\chi_{s/2}$ denote fundamental spinors.
The orbital part of the ground-state wave function is 
\begin{align}
\psi_0(\vec r) =& \;
\sqrt{ \frac{ m \sqrt{ \omega_c^2 - 2 \omega_z^2 } }{ 2 \pi } } \,
\exp\left( - \frac{m}{4} \sqrt{ \omega_c^2 - 2 \omega_z^2 } \, \rho^2 \right) 
\nonumber\\[0.1133ex]
& \; \times \left( \frac{ m \omega_z }{\pi} \right)^{1/4} \,
\exp\left( - \frac12 m \omega_z z^2 \right) \,,
\end{align}
where 
\begin{equation}
\omega_\plus - \omega_\minus = 
\sqrt{ \omega_c^2 - 2 \omega_z^2 } \,.
\end{equation}
The spin-up sublevel of the cyclotron ground state,
and the spin-down sublevel of the first excited cyclotron state,
are of interest for spectroscopy and determination
of the anomalous magnetic moment of the 
electron (see also Fig.~\ref{fig3}).
The spin-up sublevel of the cyclotron ground state 
fulfills the relations
\begin{subequations}
\label{level1}
\begin{align}
& \; H_0 \, \psi_{0 0 0 1}(\vec r) = 
E_{0 0 0 1} \, \psi_{0 0 0 1}(\vec r) \,,
\\[0.1133ex]
& \; E_{ 0 0 0 1 } = 
\frac{\omega_c}{2} ( 1+\kappa ) +
\frac{\omega_\plus}{2} + \frac{\omega_z}{2} - \frac{\omega_\minus}{2} \,.
\end{align}
\end{subequations}
The spin-down sublevel of the first excited cyclotron state
fulfills the relations
\begin{subequations}
\label{level2}
\begin{align}
& \; H_0 \, \psi_{0 0 1 \; -1}(\vec r) = 
E_{ 0 0 1 \; -1 } \, \psi_{0 0 1 \; -1}(\vec r) \,,
\\[0.1133ex]
& \; E_{ 0 0 1 \; -1 } = - \frac{\omega_c}{2} (1+\kappa) 
+ \frac{3 \omega_\plus}{2} + \frac{\omega_z}{2} - \frac{\omega_\minus}{2} \,.
\end{align}
\end{subequations}
Here we consider, for simplicity, the sublevels with 
$k = \qq = 0$, i.e., without axial or magnetron excitations.
This approximation will be lifted in Sec.~\ref{sec4}.
Due to the anomalous magnetic moment,
$E_{0 0 0 1}$ is a little higher than
$E_{0 0 1 -1}$, and the energy difference is
\begin{align}
\Delta E =& \;
E_{0 0 0 1} - E_{0 0 1 \, -1} =
\omega_c (1+\kappa) - \omega_\plus 
\nonumber\\[0.1133ex]
=& \; \omega_c (1+\kappa) - \frac12 \, \left( \omega_c +
\sqrt{ \omega_c^2 - 2 \omega_z^2 } \right) \,.
\end{align}
In the limit $\omega_z \to 0$, one has 
$\Delta E \to \kappa \, \omega_c$,
which relates energy levels inside the trap 
to the anomalous magnetic moment of the electron.
The energy difference $\Delta E$ serves to 
determine the anomalous magnetic moment of the 
electron~\cite{BrGa1982,BrGaHeTa1985,GaEtAl2006everything,%
HaFoGa2008,FaGa2021prl,FaGa2021pra}.

%
% Scaling for Strong Fields and Penning Trap
%
\subsection{Scaling for Strong Fields and Penning Trap}
\label{sec23}

In order to illustrate the analogy of the 
bound spectrum inside a Penning trap 
and an electron bound in an atom,
we introduce coupling parameters.
With reference to the QED coupling $\alpha_\QED = 
e^2/(4 \pi) \approx 1/137.036$, we refer to these 
as the cyclotron coupling parameter
$\alpha_c$ (which could otherwise be referred to as the 
cyclotron fine-structure constant),
and the axial coupling parameter $\alpha_z$,
\begin{equation}
\label{defalpha}
\alpha_c = \sqrt{\frac{\omega_c}{m}} \,,
\qquad
\alpha_z = \sqrt{\frac{\omega_z}{m}} \,.
\end{equation}
A third coupling parameter, 
pertaining to the magnetron frequency, 
is defined in Eq.~\eqref{defalpha_m}.
Because of the hierarchy of 
typical frequencies in a trap~\cite{BrGa1986}, one has
\begin{equation}
\alpha_z \ll \alpha_c \,.
\end{equation}
Once the cyclotron and the axial frequencies are defined, 
we can calculate the magnetron frequency
based on Eq.~\eqref{defomegaminus}.
One can define a trap fine-structure constant $\alpha_\PT$ 
in terms of the maximum of the coupling parameters
$\alpha_c$ and $\alpha_z$,
\begin{equation}
\alpha_\PT = \max( \alpha_z, \alpha_c) \,,
\qquad
\omega_\PT = \max( \omega_z, \omega_c) \,.
\end{equation}
Then, we can define scaling parameters $\xi_c$ and $\xi_z$ by
\begin{equation}
\label{xicdef}
\alpha_c = \xi_c \, \alpha_\PT \,,
\qquad
\alpha_z = \xi_z \, \alpha_\PT \,,
\qquad
\max( \xi_c, \xi_z ) = 1 \,.
\end{equation}
For the magnetron coupling parameter $\alpha_\magnetron$,
it follows that
\begin{subequations}
\label{defalpha_m}
\begin{align}
\alpha_\magnetron =& \;
\sqrt{\frac{\omega_\minus}{m}} = \xi_\magnetron \, \alpha_\PT \,,
\\
\xi_\magnetron =& \; \frac{1}{\sqrt{2}}
\left( \xi_c^2 - \sqrt{ \xi_c^4 - 2 \xi_z^4 } \right)^{1/2} \approx
\frac{ \xi_z^2 }{ \sqrt{2} \, \xi_c } \,,
\end{align}
\end{subequations}
where $\xi_\magnetron$ is smaller than either 
$\xi_c$ or $\xi_z$.
Electron momenta in the trap can be shown to be of order
\begin{equation}
p_\PT \sim \alpha_\PT \, m \,,
\end{equation}
in analogy to an atom, where $\alpha_\PT$ would be 
replaced by $\alpha_\QED$.
(By $\sim$ we indicate that
the quantities on the right and left are
of the same order-of-magnitude, while 
$\approx$ is reserved to indicate approximate equality.)
In atoms, the wave function is spread over a length scale
commensurate with the Bohr radius
$a_0 = \hbar/p$, where $p$ is a characteristic momentum.
We conclude that the ``trapped Bohr radius'' $a_{0 \PT}$ is
\begin{equation}
a_{0 \PT} = \sqrt{ \frac{ \hbar }{ m \, \omega_\PT } } =
\frac{1}{\alpha_\PT \, m} \sim \frac{ \hbar }{ p_\PT }\,.
\end{equation}
With these definitions, we can establish the scaling
of frequencies, momenta and position operators inside a
trap. In view of Eqs.~\eqref{defomegac} and~\eqref{defalpha}, 
we have
\begin{equation}
\omega_c = \frac{|e| \, B_\PT}{m} 
= \alpha_c^2 \, m \sim \alpha_\PT^2 \, m \,.
\end{equation}
It is clear that the position vector $\vec r$ scales as
\begin{equation}
| \vec r | \sim a_{0 \PT} = \frac{1}{\alpha_\PT m} \,.
\end{equation}
The scaling of the quadrupole potential follows as
\begin{equation}
V \sim \omega_z^2 (z^2 - \rho^2) \sim \alpha_\PT^4 \; \alpha_\PT^{-2} m 
= \alpha_\PT^2 m \,,
\end{equation}
and also
\begin{equation}
e \vec A_T \sim e B_\PT |\vec r| \sim 
\alpha_\PT^2 \, \alpha_\PT^{-1} \, m =
\alpha_\PT \, m \,.
\end{equation}
Finally, the kinetic momentum in the trap,
defined in Eq.~\eqref{defpiT}, isa  of the order of
\begin{equation}
\vec \pi_\PT = \underbrace{ \vec p }_{ \sim \alpha_\PT \, m } - 
\underbrace{ e \vec A_\PT }_{ \sim \alpha_\PT \, m } 
\sim \alpha_\PT \, m \,.
\end{equation}
So, the appropriate scaling for the trap implies the 
following relations, which we summarize for convenience.
\begin{subequations}
\begin{align}
\label{scaletrap}
\vec \pi \sim & \; \alpha_\PT \,, \qquad
e \vec A_\PT \sim \alpha_\PT \,, \qquad
e \, \vec B_\PT \sim \alpha_\PT^2 \,, 
\\[0.1133ex]
e \, \vec E_\PT =& \; -\vec\nabla V \sim \alpha_\PT^3  \,,
\qquad
e \, \partial_t \vec E \sim \alpha_\PT^5 \,.
\end{align}
\end{subequations}
The second of these implies that, if we wish to 
calculate Penning trap energy levels to 
order $\alpha_\PT^2$, then we need to keep all terms
quadratic in the magnetic trap field,
and if we wish to calculate them to order $\alpha_\PT^6$, then
we need to keep all terms cubic in the magnetic fields.
The scaling with the coupling parameters is 
notably different from atomic systems~\cite{BeSa1957}.
An expansion to third order in the magnetic fields
is not necessary for atoms, where terms 
of higher than the second order in the magnetic 
fields can be safely 
discarded~\cite{Pa2005,ZaPa2010,PaYePa2016,HiLePaSo2013,HaZhKoKa2020}.
This necessity, in addition to the other aspects 
described in Sec.~\ref{sec1}, motivates revisiting 
the Foldy--Wouthuysen transformation for general fields.
The scaling of the magnetic field in a Penning trap 
is completely different from that in an atom.
In typical atoms, the only important magnetic field is 
the dipole magnetic field generated 
by the atomic nucleus, which leads to the hyperfine splitting.
In the Penning trap, by contrast, the magnetic field 
provides for the binding of the electron,
which is why it needs to be taken out in higher orders.

%
% Time Derivative and Foldy--Wouthuysen
%
\subsection{Time Derivative and Foldy--Wouthuysen}
\label{sec24}

Let us briefly review the formalism of the 
Foldy--Wouthuysen transformation, with a particular 
emphasis on the time derivative term.
The Foldy--Wouthuysen method~\cite{FoWu1950} is based on 
a unitary transformation
\begin{equation}
U = \exp(\ii S)  \,.
\end{equation}
In order to consistently derive the formalism,
it is necessary to realize that 
the time derivative of an operator does not necessarily 
commute with the operator itself.
The transformation is constructed so that,
iteratively, the Foldy--Wouthuysen
transformed Dirac Hamiltonian 
is given as follows,
\begin{align}
\label{cagancho}
\calH_{\rm FW} = & \; \exp(\ii S) \, 
\left[ \calH - \ii \partial_t \right] \,  \exp(-\ii S) 
\nonumber\\
=& \; \calH + [ \ii S, \calH - \ii \partial_t ] 
+ \frac{1}{2!} \,  [ \ii S,  [ \ii S, \calH - \ii \partial_t ]  ]
\nonumber\\
& \; 
+ \frac{1}{3!} \,  [ \ii S,  [ \ii S,  [ \ii S, \calH
-\ii \partial_t ]  ] ]
+ \ldots 
\nonumber\\
=& \; 
\calH + 
\delta \calH^{(1)} + 
\delta \calH^{(2)} + 
\delta \calH^{(3)} + \ldots 
\end{align}
Here, the differential operator $\ii \partial_t$ 
is understood to exclusively act on the unitary operator $\exp(-\ii S)$,
but not on the wave function.
The time derivatives add additional terms,
which, in first order, read as follows,
\begin{equation}
\label{example}
\delta \calH^{(1)} = [ \ii S, \calH - \ii \partial_t ] = 
\ii [S, \calH ] - \partial_t S \,.
\end{equation}
Note that one can iteratively 
calculate the multi-commutators in Eq.~\eqref{cagancho},
\begin{equation}
\delta \calH^{(n+1)} = 
\frac{1}{n+1} \; [ \ii S, \delta \calH^{(n)} ]  \,.
\end{equation}

For a typical generalized Dirac Hamiltonian, 
the Foldy--Wouthuysen transformation operator $S$ is 
proportional to 
\begin{equation}
S \sim \vec\alpha \cdot \vec \pi \sim \alpha_\PT \,,
\qquad
\partial_t S \sim \vec\alpha \cdot \partial_t \vec \pi \sim \alpha_\PT^3 \,.
\end{equation}
Here, the $\vec \alpha$ and $\beta$ matrices are used in the 
Dirac representation,
\begin{equation}
\label{diracrep}
\vec\alpha = \left( \begin{array}{cc}
0 & \vec\sigma \\ \vec\sigma & 0 \\
\end{array} \right) \,,
\qquad
\beta =
\left( \begin{array}{cc}
\mathbbm{1}_{2 \times 2} & 0 \\
0 & -\mathbbm{1}_{2 \times 2}
\end{array} \right) \,,
\end{equation}
where $\vec\sigma$ denotes the vector of $2 \times 2$ Pauli matrices.
It becomes clear that the higher-order terms in the 
multi-commutator expansion~\eqref{cagancho} 
represent higher orders in the momenta.

In the Coulomb gauge, one can separate the electric field
into its longitudinal and transverse components,
\begin{equation}
\label{given}
e \vec E_\parallel = - \vec\nabla V \,,
\qquad
\vec E_\perp = - \partial_t \vec A \,.
\end{equation}
The longitudinal component is obtained as the 
commutator of kinetic momentum and potential $V$,
\begin{equation}
[\vec \pi, V ] = - \ii \, \vec\nabla V
= - \ii \, e \, \vec\nabla A^0
= \ii \, e \, \vec E_\parallel \,,
\end{equation}
while the transverse component is obtained
via the time derivative of the vector potential,
\begin{equation}
\partial_t \vec \pi = -e \, \partial_t \vec A
= e \, \vec E_\perp \,.
\end{equation}
The time derivative term
in Eq.~\eqref{example} is decisive in ensuring that the
Foldy--Wouthuysen transformed Dirac Hamiltonian
contains the complete electric field.

%
% Higher--Order Foldy--Wouthuysen Transformation
%
\section{Higher--Order Foldy--Wouthuysen Transformation}
\label{sec3}

%
% Higher--Order Corrections for Strong Fields
%
\subsection{Higher--Order Corrections for Strong Fields}
\label{sec31}

We start from the generalized 
Dirac Hamiltonian (see Chap.~7 of Ref.~\cite{ItZu1980}
and Chap.~1 of Ref.~\cite{JeAd2022book}),
including the anomalous-magnetic-moment terms.
The $g$ factor of the electron is expressed
as $g = 2 (1+\kappa)$. The Hamiltonian is
(see Chap.~7 of Ref.~\cite{ItZu1980})
\begin{equation}
\label{Hgen}
\calH = \vec\alpha \cdot \vec \pi + \beta m + V
+ \frac{\kappa e}{2 m} \left( \ii \vec\gamma \cdot \vec E - 
\beta \, \vec\Sigma \cdot \vec B \right) \,.
\end{equation}
The vector of Dirac $\gamma$ matrices is used 
as $\vec\gamma = \beta \, \vec\alpha$,
with reference to Eq.~\eqref{diracrep}.
The $\gamma$ matrices and the 
$4 \times 4$ spin matrices $\vec\Sigma$, which we
will need in the following, read as
\begin{equation}
\vec\gamma = \left( \begin{array}{cc}
0 & \vec\sigma \\ -\vec\sigma & 0 \\
\end{array} \right) \,,
\qquad
\vec\Sigma = \left( \begin{array}{cc}
\vec\sigma & 0 \\ 0 & \vec\sigma 
\end{array} \right) \,.
\end{equation}

As anticipated in Sec.~\ref{sec1}, the
iterated Foldy--Wouthuysen method aims to 
eliminate the ``odd'' operators (in bispinor
space) from the Dirac Hamiltonian, in successive
higher orders of the momenta.
If $\calO$ is the odd operator in the 
Dirac Hamiltonian, then the unitary transformation
is $U = \exp(\ii S)$ where $S = -\ii \beta \calO / (2 m)$.
For the general Hamiltonian given in Eq.~\eqref{Hgen},
one needs to employ three transformations, 
$S = S^{(1)}$, $S = S^{(2)}$ and $S = S^{(3)}$,
respectively, which are given as follows.
For the first transformation, one can easily derive 
the expression for $S^{(1)}$ from Eq.~\eqref{Hgen},
\begin{equation}
S^{(1)} = -\ii \frac{\vec \gamma \cdot \vec \pi}{2 m}
+ \frac{ e \, \kappa }{4 m^2} \vec \alpha \cdot \vec E \,.
\end{equation}
One calculates the multi-commutators given in Eq.~\eqref{cagancho},
up to seventh order.
The second transformation is used to eliminate
further remaining odd operators in the 
first transformed Hamiltonian~\cite{JeNo2013pra,Wo1999,JeNo2014jpa}.
It is more complicated,
\begin{align}
S^{(2)} = & \;
\frac{ \ii }{6 m^3} (\vec \gamma \cdot \vec \pi)^3
+ \frac{ e }{4 m^2} (\vec \alpha \cdot \vec E)
\nonumber\\
& \; 
- \frac{ \ii e \kappa}{8 m^3} [ \vec \gamma \cdot \vec \pi, \vec \Sigma \cdot \vec B ]
- \frac{ \ii }{60 m^5} \, (\vec \gamma \cdot \vec \pi)^5
\nonumber\\
& \; 
+ \frac{ e }{96 m^4} \beta \, 
[ [ \vec \gamma \cdot \vec \pi, \vec \Sigma \cdot \vec E ], \vec \Sigma \cdot \vec \pi ]
\nonumber\\
& \; 
+ \frac{ e \kappa}{12 m^4} \beta \, 
[ [ \vec \gamma \cdot \vec \pi, \vec \Sigma \cdot \vec E], \vec \Sigma \cdot \vec \pi ]
\nonumber\\
& \;
- \frac{ e \kappa}{4 m^4} \beta \,
(\vec \gamma \cdot \vec \pi) (\vec \Sigma \cdot \vec E) (\vec \Sigma \cdot \vec \pi)
\nonumber\\
& \; 
+ \frac{ \ii e \kappa}{192 m^5} 
[ \vec \gamma \cdot \vec \pi, [ \vec \Sigma \cdot \vec \pi, 
\{ \vec \Sigma \cdot \vec \pi, \vec \Sigma \cdot \vec B \} ] ] 
\nonumber\\
& \;
+ \frac{ \ii e \kappa}{48 m^5}
\vec \gamma \cdot \vec \pi \, 
\{ \vec \Sigma \cdot \vec \pi, \vec \Sigma \cdot \vec B \} \, \vec \Sigma \cdot \vec \pi
\nonumber\\
& \; 
+ \frac{ \ii e \, \kappa }{8 m^3} \vec \gamma \cdot \partial_t \vec E \,.
\end{align}
Finally, the third transformation, which 
eliminates all odd operators up to seventh order, is 
given as follows,
\begin{align}
S^{(3)} = & \;
- \frac{ \ii }{12 m^5} (\vec \gamma \cdot \vec \pi)^5
+ \frac{ \ii e }{8 m^3} (\vec \gamma \cdot \partial_t \vec E)
\nonumber\\
& \;
+ \frac{ 5 e \beta }{96 m^4} 
[ [ \vec \gamma \cdot \vec \pi, \vec \Sigma \cdot \vec E ], 
\vec \Sigma \cdot \vec \pi ] 
\nonumber\\
& \;
- \frac{ 3 e \beta }{32 m^4} 
\{ \{ \vec \gamma \cdot \vec \pi, \vec \Sigma \cdot \vec E \}, 
\vec \Sigma \cdot \vec \pi \}
\nonumber\\
& \;
+ \frac{ e^2 \kappa }{8 m^4} \beta \,
\{ \vec\gamma \cdot \vec B, \vec \Sigma \cdot \vec E \}
\nonumber\\
& \;
+ \frac{ e \kappa}{16 m^4} \beta \,
\{ \vec \Sigma \cdot \partial_t \vec B, \vec \Sigma \cdot \vec \pi \}
\nonumber\\
& \;
+ \frac{ 5 \ii e \kappa}{96 m^5} 
\{ \{ \vec \gamma \cdot \vec \pi, \vec \Sigma \cdot \vec B \}, 
(\vec \Sigma \cdot \vec \pi)^2 \}
\nonumber\\
& \;
- \frac{ \ii e \kappa}{48 m^5} 
[ [ \vec \gamma \cdot \vec \pi, \vec \Sigma \cdot \vec B ],
(\vec \Sigma \cdot \vec \pi)^2 ] \,.
\end{align}
The result of the iterative 
seventh-order standard Foldy--Wouthuysen 
transformation~\cite{JeNo2014jpa}
can be written as
\begin{equation}
\calH_{\rm FW} = \calH^{[0]} + \calH^{[2]} + \calH^{[3]} + \calH^{[4]}
+ \calH^{[5]} + \calH^{[6]} + \calH^{[7]} \,.
\end{equation}
The superscript denotes the power of the coupling parameter
at which the term becomes relevant.
The coupling parameter can either be 
$\alpha_\PT$ (for the Penning trap), or 
$\alpha = \alpha_\QED \approx 1/137.036$ (for an atom).
In zeroth order in $\alpha$, we only have the rest mass term,
\begin{equation}
\label{calH0}
\calH^{[0]} = \beta m \,,
\end{equation}
In the second order in $\alpha$, we have the nonrelativistic term
\begin{equation}
\label{calH2}
\calH^{[2]} = \beta \frac{1}{2 m} ( \vec\Sigma \cdot \vec\pi )^2 + V \,.
\end{equation}
In the third order in $\alpha$, we only have a single term,
\begin{equation}
\label{calH3}
\calH^{[3]} = - \frac{e \, \kappa}{2 m} \, \beta \, \vec\Sigma \cdot \vec B \,.
\end{equation}
where the one-loop (Schwinger) correction~\cite{Sc1948} to the 
anomalous magnetic moment of the electron is 
$\kappa = \alpha/(2 \pi)$.
The $\alpha^4$ terms can be expressed very succinctly,
\begin{equation}
\label{calH4}
\calH^{[4]} = 
-\beta \, \frac{1}{8 m^3} ( \vec\Sigma \cdot \vec\pi )^4
- \frac{\ii \, e}{8 m^2} [ \, \vec\Sigma \cdot \vec\pi,  \vec\Sigma \cdot \vec E \, ] \,.
\end{equation}
The $\alpha^5$ anomalous magnetic-moment terms are
also expressed in quite a compact form,
\begin{equation}
\label{calH5}
\calH^{[5]} = 
- \frac{\ii \, e \, \kappa}{4 m^2} \; 
[ \, \vec\Sigma \cdot \vec \pi, \vec\Sigma \cdot \vec E \, ]
+ \beta \; \frac{e \, \kappa}{16 m^3} \; 
\{ \, \vec\Sigma \cdot \vec \pi, 
\{ \, \vec\Sigma \cdot \vec \pi, 
\vec\Sigma \cdot \vec B \, \} \} \,.
\end{equation}
From the direct iterative application of the 
multi-commutator expansion~\eqref{cagancho},
one obtains the $\alpha^6$ terms,
\begin{align}
\label{calH6initial}
\calH^{[6]} =& \; 
\beta \, \frac{1}{16 m^5} ( \vec\Sigma \cdot \vec\pi )^6
\\[0.1133ex]
& \; - \frac{5 \ii e}{128 m^4} 
\underbrace{ [ \, \vec\Sigma \cdot \vec\pi, [ \, \vec\Sigma \cdot \vec\pi, 
[ \, \vec\Sigma \cdot \vec\pi, \vec\Sigma \cdot \vec E \, ] \, ] \, ] }_%
{\equiv X}
\nonumber\\[0.1133ex]
& \; + \frac{\ii \, e }{8 m^4} \; 
\{ \, (\vec\Sigma \cdot \vec \pi)^2, 
[ \, \vec\Sigma \cdot \vec \pi, \vec\Sigma \cdot \vec E \, ] \}
+ \beta \, \frac{e^2 \vec E^{\,2} }{8 m^3} \,,
\nonumber
\end{align}
where we implicitly define the $X$ term.  
It is computationally advantageous,
in the consideration of the $\alpha^6$ terms,
to map the algebra of the commutators of the operators
onto a computer symbolic program~\cite{Wo1999}.
It is also instructive to 
present an alternative expression for the sixth-order terms $\calH^{[6]}$.
One derives the identity
\begin{align}
\label{rel1}
X =& \;
\left\{ \left( \vec\Sigma \cdot \vec\pi \right)^2,
\left[ \vec\Sigma \cdot \vec\pi, \vec\Sigma \cdot \vec E \right] \right\}
\nonumber\\[0.1133ex]
& \; - 2\, \underbrace{ \left( \vec\Sigma \cdot \vec\pi \right) \,
\left[ \vec\Sigma \cdot \vec\pi, \vec\Sigma \cdot \vec E \right] \,
\left( \vec\Sigma \cdot \vec\pi \right) }_{\equiv Y} \,.
\end{align}
The second term in the above expression can be 
reformulated as follows,
\begin{align}
\label{rel2}
Y =& \; 
\frac12 \left[ \left( \vec\Sigma \cdot \vec\pi \right)^2 ,
\{ \vec\Sigma \cdot \vec E,
\vec\Sigma \cdot \vec\pi \} \right] 
\nonumber\\[0.1133ex]
& \; - \frac12 \left\{
( \vec\Sigma \cdot \vec\pi )^2 ,
[ \vec\Sigma \cdot \vec \pi,
\vec\Sigma \cdot \vec E ] \right\} \,.
\end{align}
Using Eqs.~\eqref{rel1} and~\eqref{rel2},
we can establish that
\begin{align}
X =& \; 2 \left\{ \left( \vec\Sigma \cdot \vec\pi \right)^2,
\left[ \vec\Sigma \cdot \vec\pi, \vec\Sigma \cdot \vec E \right] \right\}
\nonumber\\[0.1133ex]
& \; - \left[ \left( \vec\Sigma \cdot \vec\pi \right)^2 ,
\left\{ \vec\Sigma \cdot \vec E, \vec\Sigma \cdot \vec\pi \right\} \right] \,.
\end{align}
The result for the sixth-order terms can thus alternatively 
be written as
\begin{align}
\label{calH6}
\calH^{[6]} =& \;
\beta \, 
\frac{ ( \vec\Sigma \cdot \vec\pi )^6 }{16 m^5} 
+  \frac{5 \ii e}{128 m^4}
\left[ \left( \vec\Sigma \cdot \vec\pi \right)^2 ,
\left\{ \vec\Sigma \cdot \vec E, \vec\Sigma \cdot \vec\pi \right\} \right] 
\nonumber\\[0.1133ex]
& \; + \frac{3 \ii \, e}{64 m^4} \;
\{ \, (\vec\Sigma \cdot \vec \pi)^2,
[ \, \vec\Sigma \cdot \vec \pi, \vec\Sigma \cdot \vec E \, ] \}
+ \beta \, \frac{e^2 \vec E^{\,2} }{8 m^3} \,.
\end{align}
The $\alpha^7$ terms contain the anomalous magnetic moment,
\begin{align}
\label{calH7}
\calH^{[7]} = & \;
\beta \, \frac{e^2 \, \kappa}{8 m^3} \, \vec E^{\,2}
- \frac{e \, \kappa}{16 m^3} \, \beta \; \{ \, \vec\Sigma \cdot \vec \pi, 
\vec\Sigma \cdot \partial_t \vec E \,  \}
\nonumber\\
& \;
+ \frac{\ii \, e \, \kappa}{16 m^4} \; 
\{ \, (\vec\Sigma \cdot \vec \pi)^2, 
[ \, \vec\Sigma \cdot \vec \pi, \vec\Sigma \cdot \vec E \, ] \}
\nonumber\\
& \;
- \beta \, \frac{e \, \kappa}{32 m^5} \; 
[ \vec\Sigma \cdot \vec \pi,
[ \vec\Sigma \cdot \vec \pi,
\{ \, \vec\Sigma \cdot \vec \pi, 
\{ \, \vec\Sigma \cdot \vec \pi, \vec\Sigma \cdot \vec B \, \} \}  \, ]]
\nonumber\\
& \;
- \beta \frac{3 e \kappa}{256 m^5} 
\{ \vec\Sigma \cdot \vec \pi,
\{ \vec\Sigma \cdot \vec \pi,
\{ \vec\Sigma \cdot \vec \pi, 
\{ \vec\Sigma \cdot \vec \pi, \vec\Sigma \cdot \vec B \, \} \}  \} \}
\nonumber\\
& \; + \frac{\ii \, e^2 \, \kappa}{16 \, m^4} \; 
[ \, \vec\Sigma \cdot \vec E, \{ \vec\Sigma \cdot \vec \pi, 
\vec\Sigma \cdot \vec B \} \, ] \,.
\end{align}
We here include all terms relevant for strong magnetic fields.

%
% General Particle Hamiltonians
%
\subsection{Particles and Antiparticles}
\label{sec32}

Let us concentrate on the upper left 
$2 \times 2$ submatrix of $\calH_{\rm FW}$, 
which is the particle (as opposed to the 
antiparticle) Hamiltonian.
It is well known that the Dirac Hamiltonian
describes particle and antiparticle states 
simultaneously~\cite{JeNo2013pra},
and that the lower right $2 \times 2$ submatrix of 
$\calH_{\rm FW}$ describes the antiparticle.
In principle, the particle Hamiltonian can be 
obtained from the results given in Eqs.~\eqref{calH2}---\eqref{calH7}
by simply replacing $\vec\Sigma \to \vec\sigma$,
and $\beta \to \mathbbm{1}_{2 \times 2}$,
but it is still instructive to give the results 
separately.

Because the rest mass term $\beta \, m$ given in Eq.~\eqref{calH0} 
is a physically irrelevant constant, we write the 
general particle Hamiltonian $H$ under the presence of the 
external electric and magnetic fields as
\begin{equation}
\label{genpart}
H = H^{[2]} + H^{[3]} + H^{[4]} + H^{[5]} + H^{[6]} + H^{[7]} \,,
\end{equation}
where we take into account up to seventh-order terms.
One finds
\begin{equation}
\label{H2}
H^{[2]} + H^{[3]} 
=  \frac{\vec\pi^{\,2}}{2 m} + V 
- \frac{e \, (1 + \kappa)}{2 m} \,  \, \vec\sigma \cdot \vec B \,,
\end{equation}
where we have used the identity
$( \vec\sigma \cdot \vec\pi )^2 =
\vec\pi^{\, 2} - e \vec\sigma \cdot \vec B$.
The fourth-order terms find the compact representation
\begin{equation}
\label{H4}
H^{[4]} = 
- \, \frac{1}{8 m^3} ( \vec\sigma \cdot \vec\pi )^4
- \frac{\ii \, e}{8 m^2} [ \, \vec\sigma \cdot \vec\pi,  \vec\sigma \cdot \vec E \, ] \,,
\end{equation}
while the fifth-order terms contain 
the anomalous magnetic-moment,
\begin{equation}
\label{H5}
H^{[5]} = 
- \frac{\ii \, e \, \kappa}{4 m^2} \; 
[ \, \vec\sigma \cdot \vec \pi, \vec\sigma \cdot \vec E \, ]
+  \; \frac{e \, \kappa}{16 m^3} \; 
\{ \, \vec\sigma \cdot \vec \pi, 
\{ \, \vec\sigma \cdot \vec \pi, 
\vec\sigma \cdot \vec B \, \} \} \,.
\end{equation}
The general $\alpha^6$ terms are given as
\begin{align}
\label{H6}
H^{[6]} =& \;
 \, \frac{ ( \vec\sigma \cdot \vec\pi )^6 }{16 m^3} 
+  \frac{5 \ii e}{128 m^4}
\left[ \left( \vec\sigma \cdot \vec\pi \right)^2 ,
\left\{ \vec\sigma \cdot \vec E, \vec\sigma \cdot \vec\pi \right\} \right] 
\nonumber\\[0.1133ex]
& \; + \frac{3 \ii \, e}{64 m^3} \;
\{ \, (\vec\sigma \cdot \vec \pi)^2,
[ \, \vec\sigma \cdot \vec \pi, \vec\sigma \cdot \vec E \, ] \}
+  \, \frac{e^2 \vec E^{\,2} }{8 m^2} \,.
\end{align}
In the above form, the sixth-order terms in the Hamiltonian are
compatible with those used 
in Eqs.~(36)--(38) of Ref.~\cite{ZaPa2010}
for spin-$1/2$ particles.
The $\alpha^6$ terms listed in Eq.~\eqref{calH6} are also
equal to those obtained by applying the unitary
transformation outlined in Eq.~(19) of Ref.~\cite{PaYePa2016}
to the Hamiltonian given in Eq.~(15) of Ref.~\cite{PaYePa2016},
i.e., to the Hamiltonian obtained by adding the
terms given in Eqs.~(15) and~(20) of Ref.~\cite{PaYePa2016}.
Also, the result in Eq.~\eqref{calH6} is equal to the Hamiltonian
considered in Eq.~(7) of Ref.~\cite{HaZhKoKa2020},
which in turn has been derived from the
NRQED approach outlined in Ref.~\cite{HiLePaSo2013}.
The $\alpha^7$ terms attain the following structure,
\begin{align}
\label{H7}
H^{[7]} = & \;
 \, \frac{e^2 \, \kappa}{8 m^3} \, \vec E^{\,2}
- \frac{e \, \kappa}{16 m^3} \,  \; \{ \, \vec\sigma \cdot \vec \pi, 
\vec\sigma \cdot \partial_t \vec E \,  \}
\nonumber\\
& \; + \frac{\ii \, e \, \kappa}{16 m^4} \; 
\{ \, (\vec\sigma \cdot \vec \pi)^2, 
[ \, \vec\sigma \cdot \vec \pi, \vec\sigma \cdot \vec E \, ] \}
\nonumber\\
& \;
-  \, \frac{e \, \kappa}{32 m^5} \; 
[ \vec\sigma \cdot \vec \pi,
[ \vec\sigma \cdot \vec \pi,
\{ \, \vec\sigma \cdot \vec \pi, 
\{ \, \vec\sigma \cdot \vec \pi, \vec\sigma \cdot \vec B \, \} \}  \, ]]
\nonumber\\
& \; -  \, \frac{3 e \, \kappa}{256 m^5} \; 
\{ \vec\sigma \cdot \vec \pi,
\{ \vec\sigma \cdot \vec \pi,
\{ \, \vec\sigma \cdot \vec \pi, 
\{ \, \vec\sigma \cdot \vec \pi, \vec\sigma \cdot \vec B \, \} \}  \,  \} \}
\nonumber\\
& \; + \frac{\ii \, e^2 \, \kappa}{16 \, m^4} \; 
[ \, \vec\sigma \cdot \vec E, \{ \vec\sigma \cdot \vec \pi, 
\vec\sigma \cdot \vec B \} \, ] \,.
\end{align}
In general, we have not found ways to simplify the 
$\alpha^7$ terms further than the expression given by Eq.~\eqref{H7}.

%
% Quantum Cyclotron Energy Levels
%
\section{Quantum Cyclotron Energy Levels}
\label{sec4}

%
% Leading Terms
%
\subsection{Leading Term}

Now, we return to the problem considered in Sec.~\ref{sec21} 
and perform simplifications for a Penning trap configuration.
The trap field $\vec B_\PT$ is assumed to be directed along the 
$z$ axis, as a constant, uniform field, so that
\begin{equation}
\label{pi2}
\vec \pi = \vec\pi_\PT = \vec p 
- \frac{e}{2} ( \vec B_\PT \times \vec r ) \,,
\quad
\vec \pi_\PT^{\,2} = \vec p^{\,2} - e \vec L \cdot \vec B_\PT + 
\frac14 m^2 \omega_c^2 \rho^2 \,,
\end{equation}
where we note the identity ($\vec B_\PT \times \vec r)^2 = B_\PT^2 \, \rho^2$.
Furthermore, we ignore the radiative (transverse) electric field and set
\begin{equation}
e \vec E = e \vec E_\parallel = - \vec \nabla V \,.
\end{equation}
Let us specialize the terms $H^{[k]}$ 
with $k=2,\dots,6$, discussed in Sec.~\ref{sec32},
to the case of a Penning trap.
The sum of the terms $H^{[2]}$ and $H^{[3]}$
is just the Hamiltonian $H_0$ given in 
Eqs.~\eqref{H0rep1} and~\eqref{H0rep2},
\begin{align}
\label{H0recall}
H_0 =& \; H^{[2]} + H^{[3]}
\\[0.1133ex]
=& \; \frac{\vec p^{\,2}}{2m}
- \frac{e}{2m} \vec L \cdot \vec B_\PT
+ \frac{m \omega_c^2}{8} \,\rho^2 + V
- \frac{e (1 + \kappa) }{2m}
\vec\sigma\cdot \vec B_\PT \,.
\nonumber
\end{align}
We recall, from Eq.~\eqref{Esnkq}, the unperturbed energy $E^{[2+3]}$ of the 
unperturbed level (eigenket $| k \qq n s \rangle$) as
\begin{align}
E^{[2+3]} =& \; E_{k \qq n s} =
\langle k \qq n s | H_0 | k \qq n s \rangle = \langle H^{[2]} + H^{[3]} \rangle
\nonumber\\[0.1133ex]
=& \; \omega_c (1+\kappa) \, \frac{s}{2}
+ \omega_\plus \left( n + \frac12 \right) 
\nonumber\\[0.1133ex]
& \; + \omega_z \left( k + \frac12 \right)
- \omega_\minus \left( \qq + \frac12 \right) \,.
\end{align}
The energy $E^{[2+3]} = E_0$ is the unperturbed (nonrelativistic)
energy.

%
% Relativistic Corrections
%
\subsection{Relativistic Corrections}

In a Penning trap,
the expression for $H^{[4]}$ simplifies as follows,
\begin{equation}
H^{[4]} =
- \frac{(\vec\sigma \cdot \vec \pi_\PT)^{\,4}}{8 m^3}
+ \frac{ \vec\nabla^2 V }{8 m^2} 
+ \frac{ \vec \sigma \, \cdot \,
(\vec\nabla V \times \vec \pi_\PT) }{4 m^2}  \,.
\end{equation}
It is adequate to treat $H^{[4]}$ and $H^{[5]}$ together.
For the fifth-order Hamiltonian $H^{[5]}$, we need the 
following relation, which is valid for a constant, 
uniform magnetic field,
\begin{multline}
\{ \, \vec\sigma \cdot \vec \pi_\PT,
\{ \, \vec\sigma \cdot \vec \pi_\PT,
\vec\sigma \cdot \vec B_\PT \, \} \} =
4 (\vec\sigma \cdot \vec \pi_\PT)^2 \,
(\vec \pi_\PT \cdot \vec B_\PT)  \,.
\end{multline}
For the Penning trap, this implies that
\begin{equation}
\label{H5PT}
H^{[5]} = \frac{\kappa \vec\nabla^2 V }{4 m^2} 
+ \frac{\kappa
\vec \sigma \, \cdot \, (\vec\nabla V \times \vec \pi_\PT)
}{2 m^2} \;
+ \frac{e \kappa \vec\sigma \cdot \vec \pi_\PT
\vec B_\PT \cdot \vec \pi_\PT }{4 m^3} \,.
\end{equation}
It is convenient to express
the sum of $H^{[4]}$ and $H^{[5]}$ as follows,
\begin{multline}
\label{H4PT}
H^{[4]} + H^{[5]} = 
- \frac{(\vec\sigma \cdot \vec \pi_\PT)^{\,4}}{8 m^3}
+ (1 + 2 \kappa) 
\frac{ \vec\nabla^2 V }{8 m^2}
\\
+ (1 + 2 \kappa) 
\frac{ \vec \sigma \, \cdot \,
(\vec\nabla V \times \vec \pi_\PT) }{4 m^2} 
+ \frac{e \kappa \vec\sigma \cdot \vec \pi_\PT
\vec B_\PT \cdot \vec \pi_\PT }{4 m^3} \,.
\end{multline}
If we are in a charge-free region, then 
$\vec\nabla^2 V = 0$, and
we have three contributions, 
\begin{equation}
H^{[4]} + H^{[5]} = H_1 + H_2 + H_3
\end{equation}
The first is the relativistic correction to the 
kinetic energy,
\begin{equation}
H_1 = - \frac{(\vec\sigma \cdot \vec\pi_\PT)^4}{8 m^3} \,.
\end{equation}
It leads to an energy shift $E_1 = \langle H_1 \rangle = 
\langle k \qq n s | H_1 | k \qq n s \rangle$ 
which reads as follows,
\begin{multline}
\label{E1}
E_1 = - \frac{ \left[ 
\frac{ \omega_\plus^2 
\left(n+\tfrac12\right) + \omega_\minus^2 
\left(\qq + \tfrac12\right) }{\omega_\plus - \omega_\minus}
+ \frac{\omega_z}{2} (k+ \tfrac12)
+ \frac{\omega_c s}{2} \right]^2 
}{2m} 
\\
- \frac{\omega_z^4
\left[ (n + \tfrac12) (\qq + \tfrac12) + \tfrac14 \right]
}{4m (\omega_\plus - \omega_\minus)^2}
- \frac{ \omega_z^2 }{16 m} 
\left[ \left(k + \tfrac12\right)^2 + \tfrac34 \right].
\end{multline}
As compared to Eq.~(7.48) of Ref.~\cite{BrGa1986},
we take the opportunity to correct an apparent 
misprint,
\begin{equation}
\underbrace{ (n + \tfrac12) (\qq + \tfrac12) - \tfrac14 }_%
{\mbox{[Eq.~(7.48) of Ref.~\cite{BrGa1986}]}} \to
\underbrace{(n + \tfrac12) (\qq + \tfrac12) + \tfrac14 }_%
{\mbox{[Result obtained here]}} .
\end{equation}
The spin-orbit coupling leads to the term
\begin{equation}
H_2 = (1 + 2 \kappa) \, \frac{1}{4 m^2} \,
\vec\sigma \cdot 
( \vec \nabla V \times \vec\pi_\PT ) \,.
\end{equation}
The energy shift $E_2 = \langle H_2 \rangle
= \langle k \qq n s | H_2 | k \qq n s \rangle$ 
reads as follows,
\begin{equation}
\label{E2}
E_2 = -\frac{\omega_z^2 s (1 + 2 \kappa)}{4 m}
\frac{ \omega_\plus (n + \tfrac12) + \omega_\minus (\qq + \tfrac12) }%
{\omega_\plus - \omega_\minus} \,.
\end{equation}
Finally, there is an  additional correction, due to the 
higher-order interaction of the electron spin 
with the magnetic field, 
\begin{equation}
H_3 = \frac{e \kappa}{4 m^3} \,
( \vec\sigma \cdot \vec\pi_\PT ) \,
( \vec B_\PT \cdot \vec\pi_\PT ) \,.
\end{equation}
The corresponding energy shift
$E_3 = \langle H_3 \rangle$ is 
\begin{equation}
\label{E3}
E_3 = - \frac{s \kappa \omega_c \omega_z}{4 m} (k + \tfrac12) \,.
\end{equation}
The terms of fourth and fifth order lead to the joint correction
\begin{equation}
E^{[4+5]} = E_1 + E_2 + E_3 \,,
\end{equation}
where the energy corrections $E_1$, $E_2$
and $E_3$ are given in Eqs.~\eqref{E1},~\eqref{E2}
and~\eqref{E3}, respectively.
In terms of the coupling parameters 
defined in Sec.~\ref{sec23}, we can express 
$E^{[4+5]}$ differently,
\begin{multline}
\label{E45}
E^{[4+5]} =
\alpha_\PT^4 m \left( - \frac{(1 + s + 2n)^2}{8} \, \xi_c^4 
\right.
\nonumber\\[0.1133ex]
- \frac{(1 + 2k)\, [1 + 2n + s (1+ \kappa)]}{8} \, \xi_c^2 \, \xi_z^2
\nonumber\\[0.1133ex]
\left.
- \frac{3 + 6 k(1+k) + 4 s (1+2n) (1+2 \kappa)}{32} \, \xi_z^4
+\mathcal{O}(\xi_z^8)
\right).
\end{multline}
The term of order $\xi_z^6$ vanishes.
If the dominant (angular) frequency in the trap 
is the cyclotron frequency, then 
we have $\xi_c = 1$ according 
to Eq.~\eqref{xicdef}.

%
% Sixth--Order Corrections
%
\subsection{Sixth--Order Corrections}

For the Penning trap, the sixth-order terms assume the form
\begin{align}
H^{[6]} =& \;
\frac{( \vec\sigma \cdot \vec\pi_\PT )^6}{16 m^5} 
-  \frac{5 \ii}{128 m^4}
\left[ \left( \vec\sigma \cdot \vec\pi_\PT \right)^2 ,
\left\{ \vec\sigma \cdot \vec \nabla V, \vec\sigma \cdot \vec\pi_\PT \right\} \right] 
\nonumber\\[0.1133ex]
& \; - \frac{3 \ii}{64 m^4} \;
\{ \, (\vec\sigma \cdot \vec \pi_\PT)^2,
[ \, \vec\sigma \cdot \vec \pi_\PT, \vec\sigma \cdot \vec \nabla V \, ] \}
+ \frac{ (\vec\nabla V)^2 }{8 m^3} \,.
\end{align}
One has two further useful relations.
The first of these is
\begin{multline}
[ (\vec\sigma \cdot \vec \pi_\PT)^2,
\{ \vec\sigma\cdot \vec \nabla V, \vec\sigma \cdot \vec\pi_\PT \} ]
= \ii \left[ \vec \pi_\PT^{\,2}, \left[ \vec \pi_\PT^{\,2}, V \right] \right]
\end{multline}
and the second,
\begin{multline}
\{ 
(\vec\sigma \cdot \vec \pi_\PT)^2,
[ \vec\sigma\cdot \vec \pi_\PT, \vec\sigma \cdot \vec \nabla V ] \} \\
= -\ii \left\{ (\vec\sigma \cdot \vec \pi_\PT)^2,
\vec \nabla^2 V + 2 \vec\sigma \cdot
\vec \nabla V \times \vec \pi_\PT \right\} \,.
\end{multline}
Hence, one can express the sixth-order terms as follows,
\begin{align}
H^{[6]} =& \;
\frac{( \vec\sigma \cdot \vec\pi_\PT )^6}{16 m^4}
+ \frac{5}{128 m^4}
\left[ \vec \pi_\PT^{\,2}, \left[ \vec \pi_\PT^{\,2}, V \right] \right]
\nonumber\\[0.1133ex]
& \; - \frac{3}{64 m^4} \;
\{ \, (\vec\sigma \cdot \vec \pi_\PT)^2,
[ \, \vec\sigma \cdot \vec \pi_\PT, \vec\sigma \cdot \vec \nabla V \, ] \}
+ \frac{ (\vec\nabla V)^2 }{8 m^3} \,.
\end{align}
Overall, in the sixth order, one has two terms, which are
{\em (i)} directly due to the sixth-order Hamiltonian
and {\em (ii)} due to a second-order effect,
involving the fourth-order Hamiltonian,
\begin{multline}
E^{[6]} = \langle k \qq n s | H^{[6]} | k \qq n s \rangle 
\\
+ \langle k \qq n s | H^{[4]}  \, \left( \frac{1}{E_{0} - H_0} \right)' \,
H^{[4]} | k \qq n s \rangle \,,
\end{multline}
where $[1/(E_{0} - H_0)]'$ 
is the reduced Green function.
A remark is in order.
For a Penning trap, the unperturbed eigenstates
are separately eigenstates of the Hamiltonian $H^{[3]}$,
and so, a conceivable additional sixth-order 
term vanishes,
\begin{equation}
\left< k \qq n s \left| H^{[3]}  \, 
\left( \frac{1}{E_0 - H_0} \right)' \,
H^{[5]} \right| k \qq n s \right> = 0 \,.
\end{equation}
The exact expression for $\langle H^{[6]} \rangle$
is very lengthy; however, an expansion into the 
coupling parameters defined in Sec.~\ref{sec23}
leads to the compact expression
\begin{multline}
\label{E6_1}
\langle H^{[6]} \rangle =
\alpha_\PT^6 m \biggl[ \frac{(1 + s + 2n)^3}{16} \, \xi_c^6
\\[0.1133ex]
+ \frac{3 (1 + 2k)\, (1 + 2n + s)^3}{32} \, \xi_c^4 \, \xi_z^2
\\
+ \frac{3}{64} \bigl[ ( 5 + 6 k(1+k) ) (1+2n) 
\\
+ (5 +6k(1+k) + 8n(1+n)) s \bigr] \, \xi_c^2 \xi_z^4
\\
+ \frac{1}{128} (1 + 2 k) (13 + 10 k (1 + k) + 6 (1 + 2 n) s) \xi_z^6
+\mathcal{O}(\xi_z^8)
\biggr] \,.
\end{multline}
After expansion in $\xi_z$, the second-order shift is 
\begin{multline}
\label{E6_2}
\left< k \qq n s \left| H^{[4]}  \, \left( \frac{1}{E_0 - H_0} \right)' \,
H^{[4]} \right| k \qq n s \right> 
\\
= \alpha_\PT^6 m \biggl[ 
-\frac{1}{32} (1 + 2 k) (1 + s + 2 n (1 + n + s)) \xi_c^4 \, \xi_z^2 
\\
- \frac{1}{64} (5 + 6 k (1 + k)) (1 + 2 n + s) \, \xi_c^2 \, \xi_z^4 
\\
- \frac{1}{512} (1 + 2 k) (45 + 17 k (1 + k) + 24 (1 + 2 n) s) \, \xi_z^6
+\mathcal{O}(\xi_z^8)
\biggr] \,.
\end{multline}
The total sixth-order energy shift is obtained 
as the sum of the results given in 
Eqs.~\eqref{E6_1} and~\eqref{E6_2},
\begin{multline}
\label{E6}
E^{[6]} = \alpha_\PT^6 m \biggl[
\frac{1}{16} (1 + 2 n + s)^3 \xi_c^6 
\\
+ \frac{ (1 + 2 k) ( 10 n (1 + n + s) + 2 + 5 s + 
     3 s^2) \xi_c^4 \xi_z^2 }{32} 
\\
+ \frac{(5 + 6 k (1 + k)) (1 + 2 n + s) + 12 n (1 + n) s) \xi_c^2 \xi_z^4 }{32}
\\
+ \frac{1}{512} (1 + 2 k) (7 + 23 k (1 + k)) \xi_z^6
+\mathcal{O}(\xi_z^8)
\biggr] \,.
\end{multline}
We leave the evaluation of the seventh-order corrections for 
a Penning trap to a future investigation.
These, otherwise, also include higher-order corrections
to the Lamb shift (self-energy) of the quantum states.

%
% Conclusions
%
\section{Conclusions}
\label{sec5}

The main results of the current investigation can be 
summarized as follows.
We have carried out, in Sec.~\ref{sec31}, the 
full iterative Foldy--Wouthuysen transformation
of the single-particle Dirac Hamiltonian, coupled to general 
electromagnetic fields, 
up to seventh order in the momenta.
The results, for the $4 \times 4$ matrices
that combine the particle and antiparticle
states, are given in Eqs.~\eqref{calH0}--\eqref{calH7}.
The effective Hamiltonian for the particle 
(as opposed to the antiparticle) is given in 
Eqs.~\eqref{H2}--\eqref{H7}.
We have clarified 
that the standard Foldy--Wouthuysen
method, iteratively applied, reproduces the 
effective Hamiltonian, to order $\alpha^7$, 
which has been derived based on NRQED methods in Ref.~\cite{HiLePaSo2013}.
So, we have shown that the Hamiltonian 
first obtained by a nonstandard Foldy--Wouthuysen
transformation in Ref.~\cite{Pa2005}, 
and then augmented by an additional 
unitary transformation given in Eq.~(19) of Ref.~\cite{PaYePa2016},
is exactly equal to the gauge-covariant Hamiltonian used
recently in Eq.~(7) of Ref.~\cite{HaZhKoKa2020}.
It is instructive to recall that the 
kinetic momentum $\vec\pi$ is gauge covariant, 
but not gauge invariant, because $\vec A$ 
transforms nontrivially under gauge transformations.
As a result of the investigations reported here,
complete agreement between the various methods
has been achieved, and
the calculations have been extended to the seventh
order in $\alpha$, including effects due 
to strong magnetic fields.

Our results are valid for cases where the 
binding of the electron proceeds in strong external
fields such as those encountered in Penning traps.
In such cases, the term $|e| B_\PT/m = \omega_c$ 
(cyclotron frequency), 
where $B_\PT$ is the trap magnetic field,
is of order $\alpha_\PT^2 m$ where $\alpha_\PT$
is a suitably defined coupling constant 
for the trap [see Eq.~\eqref{defalpha}].
In a Penning trap, the magnetic field is not an external perturbation,
but provides the decisive energy scale for the 
bound states inside the trap.
After an initial discussion of 
the separation of the electron Hamiltonian 
inside the trap, carried out in Sec.~\ref{sec21},
and the discussion of scaling relations in Sec.~\ref{sec23},
we discuss the relevant expressions for 
relativistic corrections to 
electron energy levels in quantum cyclotrons 
in Sec.~\ref{sec4}.

Our results, given for the Penning trap in
Eqs.~\eqref{E1}--\eqref{E3},~\eqref{E45} and~\eqref{E6},
enable a more accurate evaluation of 
the relativistic corrections to quantum cyclotron
states, which are important for the determination
of the fine-structure 
constant~\cite{BrGa1982,BrGaHeTa1985,GaEtAl2006everything,%
HaFoGa2008,FaGa2021prl,FaGa2021pra}.
Terms of seventh order in $\alpha_\PT$ can be obtained
from Eq.~\eqref{H7} under the
substitutions $\vec\pi \to \vec \pi_\PT$,
and $e \vec E \to -\vec\nabla V$.
However, these are of the same order as
the relativistic corrections to the Lamb shift,
notably, to the relativistic Bethe logarithm 
(order $\alpha_\PT^7 \, m$).
Hence, we leave these terms for a future work.
We do not indicate them separately.

In a more general context, our calculations show that it is possible 
to generalize the standard, direct 
calculation of the Foldy--Wouthuysen transformation
to seventh order in the coupling parameters,
under the intensive use of computer algebra~\cite{Wo1999}.

\section*{Acknowledgments}

The authors acknowledge insightful 
conversations with Professor Gerald Gabrielse.
Support from the
Templeton Foundation (Fundamental Physics Block Grant,
Subaward 60049570 of Grant ID \#{}61039)
is gratefully acknowledged.


\begin{thebibliography}{10}

\bibitem{FoWu1950}
L.~L. Foldy and S.~A. Wouthuysen, {\em \relax{On the Dirac Theory of Spin $1/2$
  Particles and Its Non-Relativistic Limit}},  Phys. Rev. {\bf 78},  29--36
  (1950).

\bibitem{BjDr1964}
J.~D. Bjorken and S.~D. Drell, {\em \relax{Relativistic Quantum Mechanics}}
  (McGraw-Hill, New York, 1964).

\bibitem{Je2014pra}
U.~D. Jentschura, {\em \relax{Fine--Structure Constant for Gravitational and
  Scalar Interactions}},  Phys. Rev. A {\bf 90},  022112  (2014).

\bibitem{ItZu1980}
C. Itzykson and J.-B. Zuber, {\em \relax{Quantum Field Theory}} (McGraw-Hill,
  New York, 1980).

\bibitem{Pa2005}
K. Pachucki, {\em \relax{Higher-order effective Hamiltonian for light atomic
  systems}},  Phys. Rev. A {\bf 71},  012503  (2005).

\bibitem{PaYePa2016}
V. Patkos, V.~A. Yerokhin, and K. Pachucki, {\em \relax{Higher-order recoil
  corrections for triplet states of the helium atom}},  Phys. Rev. A {\bf 94},
  052508  (2016).

\bibitem{BrGa1982}
L.~S. Brown and G. Gabrielse, {\em \relax{Precision spectroscopy of a charged
  particle in an imperfect Penning trap}},  Phys. Rev. A {\bf 25},
  2423(R)--2425(R)  (1982).

\bibitem{Br1985aop}
L.~S. Brown and G. Gabrielse, {\em \relax{Geonium Lineshape}},  Ann. Phys.
  (N.Y.) {\bf 159},  62--98  (1985).

\bibitem{BrGa1986}
L.~S. Brown and G. Gabrielse, {\em \relax{Geonium theory: Physics of a single
  electron or ion in a Penning trap}},  Rev. Mod. Phys. {\bf 58},  233--311
  (1986).

\bibitem{ZaPa2010}
J. Zatorski and K. Pachucki, {\em \relax{Electrodynamics of finite-size
  particles with arbitrary spin}},  Phys. Rev. A {\bf 82},  052520  (2010).

\bibitem{HaZhKoKa2020}
M. Haidar, Z.-X. Zhong, V.~I. Korobov, and J.-P. Karr, {\em
  \relax{Nonrelativistic QED approach to the fine- and hyperfine-structure
  corrections of order $m \alpha^6$ and $m \alpha^6 (m/M)$: Application to the
  hydrogen atom}},  Phys. Rev. A {\bf 101},  022501  (2020).

\bibitem{HiLePaSo2013}
R.~J. Hill, G. Lee, G. Paz, and M.~P. Solon, {\em \relax{NRQED Lagrangian at
  order $1/M^4$}},  Phys. Rev. D {\bf 87},  053017  (2013).

\bibitem{Sc1948}
J. Schwinger, {\em \relax{On Quantum-Electrodynamics and the Magnetic Moment of
  the Electron}},  Phys. Rev. {\bf 73},  416--417  (1948).

\bibitem{FaGa2021prl}
X. Fan and G. Gabrielse, {\em \relax{Circumventing Detector Backaction on a
  Quantum Cyclotron}},  Phys. Rev. Lett. {\bf 126},  070402  (2021).

\bibitem{FaGa2021pra}
X. Fan and G. Gabrielse, {\em \relax{Driven one-particle quantum cyclotron}},
  Phys. Rev. A {\bf 103},  022824  (2021).

\bibitem{BrGaHeTa1985}
L.~S. Brown, G. Gabrielse, K. Helmerson, and J. Tan, {\em \relax{Cyclotron
  Motion in a Microwave Cavity: Possible Shifts of the Measured Electron $g$
  Factor}},  Phys. Rev. Lett. {\bf 55},  44--47  (1985).

\bibitem{GaEtAl2006everything}
B. Odom, D. Hanneke, B. D’Urso, and G. Gabrielse, New Measurement of the
  Electron Magnetic Moment Using a One–Electron Quantum Cyclotron, Phys. Rev.
  Lett. 97, 030801 (2006); G. Gabrielse, D. Hanneke, T. Kinoshita, M. Nio, and
  B. Odom, New Determination of the Fine Structure Constant from the Electron
  $g$ Value and QED, ibid. 97, 030802 (2006); Erratum: New Determination of the
  Fine Structure Constant from the Electron $g$ Value and QED [Phys. Rev. Lett.
  97, 030802 (2006)], 99, 039902(E) (2007).

\bibitem{HaFoGa2008}
D. Hanneke, S. Fogwell, and G. Gabrielse, {\em \relax{Measurement of the
  Electron Magnetic Moment and the Fine Structure Constant}},  Phys. Rev. Lett.
  {\bf 100},  120801  (2008).

\bibitem{BeSa1957}
H.~A. Bethe and E.~E. Salpeter, {\em \relax{Quantum Mechanics of One- and
  Two-Electron Atoms}} (Springer, Berlin, 1957).

\bibitem{JeAd2022book}
U.~D. Jentschura and G.~S. Adkins, {\em \relax{Quantum Electrodynamics: Atoms,
  Lasers and Gravity}} (World Scientific, Singapore, 2022).

\bibitem{JeNo2013pra}
U.~D. Jentschura and J.~H. Noble, {\em \relax{Nonrelativistic limit of the
  Dirac--Schwarzschild Hamiltonian: Gravitational {\em Zitterbewegung} and
  gravitational spin-orbit coupling}},  Phys. Rev. A {\bf 88},  022121  (2013).

\bibitem{Wo1999}
S. Wolfram, {\em \relax{The Mathematica Book}}, 4 ed. (Cambridge University
  Press, Cambridge, UK, 1999).

\bibitem{JeNo2014jpa}
U.~D. Jentschura and J.~H. Noble, {\em \relax{Foldy--Wouthuysen transformation,
  scalar potentials and gravity}},  J. Phys. A {\bf 47},  045402  (2014).

\end{thebibliography}
\end{document}